\documentclass[10pt]{iopart}
\usepackage{fleqn,epsfig,iopams,amssymb,amsbsy,xspace}  
\begin{document}
     \newcommand{\pathstrange}{/home/rafelski/figure/}
     \newcommand{\pathmarc}{/usr1/rafelski/figure/}
     \newcommand{\pathbook}{/usr1/rafelski/book/figures/}
     \newcommand{\pathtransp}{/usr1/rafelski/paper/transp/}
     \newcommand{\pathlaptop}{/home/rafelski/figures/}
     \newcommand{\pathlapbook}{/home/rafelski/book/qgp/figures/}
     \newcommand{\pathletes}{/users/lpthe/jletes/bookraf/figures/}
     \newcommand{\pathjussieu}{/users/visit/rafelski/figures/}
     \newcommand{\pathnow}{}
\renewcommand{\topfraction}{.99}\renewcommand{\textfraction}{.0}
\newcommand{\myfig}[7]{%
\begin{figure}[#7]
\vskip #5cm	\centerline{\hspace*{#1cm}
\epsfig{width=#2cm,angle=#4,figure=\pathnow #3.ps}
	}\vskip #6cm
                    }
\newcommand{\capt}[3]%
{\caption[{#2}]{\label{Fig:#1}#3}
}
\newcommand{\rf}[1]{Fig.\,\ref{Fig:#1}%
}
\def\qgp{quark--gluon plasma\xspace}
\def\QGP{\qgp}

\newcommand{\ress}[1]{subsection~\ref{ssec:#1}%
}
\newcommand{\ressa}[2]{subsections~\ref{ssec:#1},~\ref{ssec:#2}%
}
\newcommand{\ressc}[2]{subsections~\ref{ssec:#1}--\ref{ssec:#2}%
}
\newcommand{\ressb}[3]{section~%
\ref{ssec:#1},~\ref{ssec:#2},~\ref{ssec:#3}%
}
\newcommand{\pref}[1]{on page~\pageref{eq:#1}} 
\newcommand{\req}[1]{Eq.\,(\ref{eq:#1})%
}
\newcommand{\reqp}[1]{Eq.\,(\ref{eq:#1}) on page~\pageref{eq:#1}%
}
\newcommand{\reqa}[2]{Eqs.\,(\ref{eq:#1}, \ref{eq:#2})%
}
\newcommand{\reqc}[2]{Eqs.\,(\ref{eq:#1}--\ref{eq:#2})%
}
\newcommand{\reqb}[3]{Eqs.\,(\ref{eq:#1}, \ref{eq:#2}, \ref{eq:#3})%
}
\newcommand{\beql}[1]{
	\begin{equation} \label{eq:#1}}

\newcommand{\beqarl}[1]{
	\begin{eqnarray} \label{eq:#1} }
\newcommand{\eeql}[1]{\label{eq:#1} \end{equation} 
} 
\newcommand{\eeqarl}[1]{\label{eq:#1} \end{eqnarray} 
}
\newcommand{\rt}[1]{table~\ref{Tab:#1}%
}
\def\beq{\begin{equation}}
\def\eeq{\end{equation}}
\def\neeq{\nonumber \eeq}
\def\beqar{\begin{eqnarray}}
\def\eeqar{\end{eqnarray}}
\def\bcite{\cite}
\def\agev{{$A$~GeV}\xspace}
\def\AGeV{\agev}
\newcommand{\captt}[3]%
{\caption[{#2}]{\label{Tab:#1}#3%
}%
}
\newcommand{\lsec}[1]{\label{sec:#1} } 
\newcommand{\res}[1]{section~\ref{sec:#1}%
}
\newcommand{\resa}[2]{sections~\ref{sec:#1},~\ref{sec:#2}%
}
\newcommand{\resc}[2]{sections~\ref{sec:#1}--\ref{sec:#2}%
}
\newcommand{\resb}[3]{section~%
\ref{sec:#1},~\ref{sec:#2},~\ref{sec:#3}%
}
\newcommand{\lssec}[1]{\label{ssec:#1} 
 }

\def\etc{{etc.\xspace}}
\def\ie{{i.e.\xspace}}
\def\eg{{e.g.\xspace}}
\def\etal{{\it et al.\/}}
\def\viz{{viz.\xspace}}
\def\Jpsi{${\mathrm J}\!/\!\,\Psi$\xspace}
\newcommand{\myfigd}[9]{%
\begin{figure}[#9]
\vskip #5cm	\centerline{\hspace*{#1cm}
\psfig{width=#2cm,angle=#4,figure=\pathnow #3.ps}
\hspace*{#7cm}
\psfig{width=#2cm,angle=#4,figure=\pathnow #8.ps}
	}\vskip #6cm
                    }
\hyphenation{re-commend-ed}

\title{Strangeness, Equilibration, Hadronization}

\author{Johann Rafelski}

\address{Department of Physics,
University of Arizona, Tucson, AZ 85721\\
and\\ CERN-Theory Division, 1211 Geneva 23, Switzerland}

\begin{abstract}
In these remarks I  explain the motivation which leads us
to consider chemical nonequilibrium processes in flavor equilibration  
and  in statistical hadroniziation of  quark--gluon plasma (QGP).
Statistical hadronization allowing for chemical non-equilibrium
is introduced. The reesults of fits to 
 RHIC-130 results, including 
multistrange hadrons, are shown to agree only with the 
model of an exploding QGP fireball.
\end{abstract}


\submitto{\JPG \rm Proceedings of Strange Quark Matter 2001, Frankfurt}


\vskip -10cm \ \hfill CERN-TH/2001-364 \vskip 10cm

\section{Historical background}      
The topic we discuss today, production of hadrons in statistical
hadronization in high energy heavy ion collisions, has been the 
subject of several diploma and doctor thesis at the University 
Frankfurt in late 70's and early 80's. At the time, in
a field void of any experimental result, I competed with a Hungarian
hadrochemistry group lead by the chair of this discussion session,
Prof. J. Zimanyi.

We both passed through natural evolution stages. We 
were at first considering the equilibrium particle abundances expected
to arise when Mr.\,EquilibriX  holds in his hand
hot and dense hadron matter fireball, which evaporated these 
particles. This work was followed by the
development of kinetic theory of 
strangeness production, which was precipitated by
the finding of the Budapest group, that light quark
interactions were not fast enough to produce strangeness 
abundantly.

We learned to appreciate that the yields of newly produced quarks
were not necessarily given by the chemical equilibrium
statistical model.
The physics reason which lead us to explore the kinetic theory 
was relatively short  available reaction time  in relativistic 
heavy ion collisions, in general
not allowing for the achievement 
of an equilibrium particle abundance. These remarks apply 
to the case that the dense matter fireball is made of hadrons, 
or deconfined quarks. Our work was extended 
to compare these two so different matter phases. 

This kinetic theory  work has inspired the  chemical non-equilibrium 
study of hadronization of quark--gluon plasma (QGP), and we 
were able to show that in a sudden hadronization process 
specific properties of QGP will be imparted in hadronization on
particle yields. All these developments were then summarized in a
relatively widely known theoretical review \bcite{Koc86}. 

When the first multistrange hadron experimental results arrived
10 years ago, the analysis applied 
introduced the chemical non-equilibrium \bcite{Raf91a}, 
which then  was generally accepted 
as being an important element of data analysis. This analysis 
addressed solely the strange particle abundances. When results for 
 non-strange hadrons were included and were analyzed, it became 
necessary to expand the chemical-non-equilibrium approach, allowing the
abundances of light quarks to vary \bcite{Let99,Let99c}. In the analysis
of the 158$A$ GeV Pb collisions with Pb targets, we discovered that the
light quark overabundance converged towards the maximum value allowed, at
which point the entropy content of the final hadronic state is
maximized \bcite{Let00b}. 

In short, we expected chemical nonequilibrium, excess of 
entropy (hadron multiplicity), and 
abundant strange antibaryons. Experiments, too numerous to 
cite, but present in this volume, found these interesting signatures
of new physics, just as predicted. So why are we having this 
discussion? As we evolved in 
our detailed understanding of the subject, we could see in our 
research field the
arrival of new people who were often fast, but sometimes slow, in
 repeating our evolutionary steps. 

This is
not the place to give those  who followed the  credit for not reading 
our work. However, we need to deal with the current situation that
our early rather trivial work of 1979--81 is today taken seriously again, and is 
used as basis of data analysis and interpretation. 
Even if this happens without citation,
this is not changing the problem we are facing: if this work
is now widely reported, is this `majority' vote suggesting that the 
very extensive theoretical development carried out by many prominent 
theorists over the  past 20  years  has gone in the wrong direction? 
Clearly, this is not the case, so we have  at least to make an effort
to end this rather absurd situation.

The principal speaker (K. Redlich) has presented his views,
which we address in technical manner in our main
contribution to this volume \bcite{Raf02}. This 
contribution is a qualitative discussion of approaches and results
in the study of the experimental data.

\section{Strangeness Equilibration}      
Collision of small elementary systems produce colored partons
linked by strings, which `snap' giving birth to 
particle pairs, and strangeness is  in p--p interactions
produced 1/5 as often compared to the light flavor u or d. 
The reasons this happens are understood in the realm
of string tension and particle tunneling. Such 
results form the basis of simulations of nuclear collisions
with independent string mechanics. 

The formation of a deconfined region of space-time 
is at origin of another effective  mechanism of strangeness 
production. Beside (confined or deconfined) quarks, we now also have 
 gluons in the hot gas, and  glue--glue collisions 
are believed to be the source of the abundant strangeness
observed in the experiment. The string mechanism uses the 
chromo-electric field between color charges, the gluon is the
transverse quantum of the chromo-fields. Thus, though in both
cases mechanisms specific to the glue filed are invoked,
both mechanisms are complementary.

It is a mere coincidence that given the strength of the string
tension (1 GeV/fm), the mass of the strange quark ($160\pm50$ MeV), 
and the initial temperatures we reach  in the hot fireball (200--300 MeV), 
that in the p--p interactions
and central Pb--Pb interactions the yield of strangeness 
is near to chemical equilibrium yield. A lot of theoretical
effort went to understand this situation within kinetic 
models. This fact cannot be generalized to different
collision systems without prior and thorough study of the
kinetic production  mechanism applicable.
The extension of the work on strangeness is 
the study of light quark abundance. It is not 
guaranteed that at all conditions, and for all 
systems considered even $u,d$ quarks are found in 
chemical equilibrium.

The deviations from
chemical equilibrium  can be incorporated into
the momentum distribution of (strange) quarks and antiquarks 
in the deconfined matter in the form:
\beql{ssbardis}
f_{{\rm s}/{\overline{\rm s}}}=
  {1\over \gamma_{\rm s}^{-1}\lambda_{\rm s}^{\mp 1} e^{\sqrt{p^2+m_{\rm s}^2}/T}+1}
\quad    \to \gamma_{\rm s} \lambda_{\rm s}^{\pm 1} e^{-\sqrt{p^2+m_{\rm s}^2}/T} 
\eeq
When $\gamma_{\rm s}\to 1$ we are approaching 
chemical equilibrium distribution. On the right hand
side in \req{ssbardis}, we see the Boltzmann limit. This distribution maximizes
the entropy at given particle number. When a Boltzmann momentum
distribution is established, the phase space occupancy
$\gamma_{\rm s}(t)$ may be still quite different from 
unity, as indicated it evolves in time $t$ due to 
ongoing pair production (gluon fusion) processes, $\rm GG\to s\bar s$.

The characteristic time 
required to equilibrate strangeness chemically in 
the deconfined gluon gas 
is shown \rf{figTaussrun}. This result suggests that 
depending on the initial temperature and subsequent evolution, 
the lifespan of the QGP needs
to live at high temperature in excess of 4 fm/c in order for the 
strange quark flavor to reach full chemical equilibrium. 
This lifespan of the plasma phase is not generally assured,
and thus in QGP the chemical equilibrium is not easily reached.

\myfig{0.5}{12}{TAUQGTSRUNPARIS2R}{0}{-4.5}{-0.9}{tb}
\capt{figTaussrun}{QGP strangeness relaxation time}{
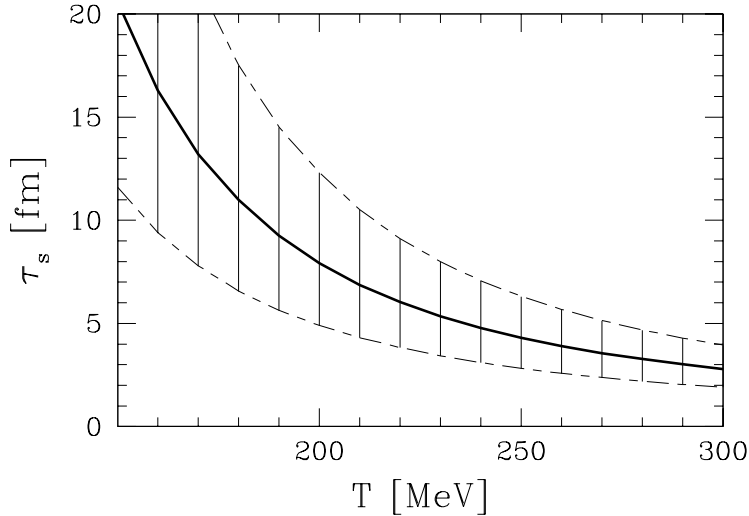
Strangeness chemical relaxation 
time $\tau_{\rm s}$, for $\alpha_{s}(M_{\rm Z})=0.118$
with $m_{\rm s}(M_{{\rm Z}})=90$~MeV,
$\rho_{\rm s}^\infty(m_{\rm s}\simeq 200\,\mbox{MeV})$ (thick line). 
Hatched areas: effect of variation of strange quark mass by 20\%.
}
\end{figure}

\section{Hadronization}      
Hadrons  are always the most abundant final particles observed 
experimentally. We must be able to understand and interpret the
production mechanism of these particles. Strange hadrons 
are the largest particle family, and are  thus
particularly interesting.  Our primary objective is to
identify what changes can be expected in the 
yields of strange particles emitted from deconfined 
quark-gluon matter. And we must device a mechanism of data
analysis that is sufficiently sensitive to recognize this.

For now nearly 50 years  we know that the 
abundances of hadrons produced in elementary and heavy
ion collisions can be, within a factor of two, described 
by the statistical phase space. 
Hagedorn \bcite{Hag65}, and before him, Fermi \bcite{Fer50},
Landau \bcite{Lan53}, have pioneered
that idea.  One can argue that when the production probability 
(the square of the quantum matrix element) 
 of different hadronic
particles is similar,  they saturate by strength of their 
interactions the unitarity limit, the primary difference 
in particle yield is arising from the relative magnitude of 
the accessible emission phase space. 

This reduces the issue to the question, why can we 
use a `temperature' $T$ parameter  to characterize
the size of hadronic particle phase space?
Hagedorn spoke of a pre-existent temperature, later
he developed a bootstrap model which helped quantify
this idea \bcite{Hag73}. Another possible approach uses
 the quark-vacuum fluctuations 
to obtain  Boltzmann-like hadron
momentum distribution \bcite{Bia99}. 
In nuclear collisions, a large number of quarks and gluons
 within a deconfined local domain of  space time
 can approach by two body collision processes the local thermal 
distribution within a relatively short period of time. 
In  both cases, the confining vacuum is the critical 
input factor, as it keeps together these strongly interacting 
quanta.  In my view, the confining vacuum makes strong
interactions different and introduces the statistical
 phase space into hadronic particle abundances. 

Two 
properties, both associated with gluons in the deconfined 
state, allow to remember  the structure
of the fireball, and eventually help us understand the properties of
QGP. The addition of  gluonic
 degrees of freedom is doubling the entropy
content of quarks alone, which already benefit from the breaking of
color bonds to increase in the deconfined state the entropy
content;  gluon  based reactions are effective in producing
 strange quarks. 
The enhancement of the entropy content is seen in
enhanced pion yields, while enhanced strangeness abundance
contributes to a pattern of strange hadron which favors production
of multistrange hadrons.

What is the level of sensitivity in particle detection
which is required for 
these observables? The entropy enhancement translates
into a 30--60\% increased pion yield.
Strangeness enhancement is expected at the level of 
factor two--three. One would imagine, seen the
magnitude of these effects of QGP, that we need to measure these
quantities at the level of a few percent. Thus, we must obtain
experimental results at this level of precision, and 
device an analysis method which is sensitive at this precision
level.

In quantitative terms, 
the relative number of  final state hadronic particles
freezing out from, {\eg}, a thermal quark--gluon source, is obtained
noting that the  fugacity $f_i$ of the $i$-th emitted  composite  
hadronic particle containing $k$-components is derived 
from fugacities $\lambda_k$ and phase space occupancies $\gamma_k$:
\beql{abund}
N_i\propto e^{-E_i/T_{\rm f}}f_i=e^{-E_i/T_{\rm f}}\prod_{k\in i}\gamma_k\lambda_k.
\end{equation}
As seen in \req{abund}, we study particle 
production in terms of five statistical parameters
$T, \lambda_{\rm q}, \lambda_{\rm s}, \gamma_{\rm q},  \gamma_{\rm s}$. 
The difference between the two  types of chemical parameters,
$\lambda_i$ and $\gamma_i$, is that the phase space
occupancy  factor $\gamma_{i}$ regulates the number of pairs of flavor `$i$', 
and hence applies in the same manner to particles and antiparticles, while
fugacity $\lambda_i$ applies only to particles, while $\lambda_i^{-1}$ is
the antiparticle fugacity, as is seen in \req{ssbardis}. 

To describe the shape of spectra, one needs  matter flow velocity
parameters, these become irrelevant when  total particle abundances
are studied, obtained integrating  all of (the velocity deformed)
phase space. In presence of strong longitudinal flow, the `scaling' 
in rapidity  implies that when we are looking at a
yield per unit of rapidity, we can also
ignore the velocity parameter when considering particle ratios.

The resulting yields of final state hadronic particles are most 
conveniently characterized taking the Laplace transform of the 
accessible phase space. This approach generates a  function which,
in its mathematical properties, is identical to 
the hadron gas partition function:
\beql{4abis}
{\cal L}\left[e^{-E_i/T_{\rm f}}\prod_{k\in i}\gamma_k\lambda_k\right] 
 \propto \ln{\cal Z}^{\rm HG}\,.
\eeq
This characterization of the accessible phase space
 does not require formation of a phase comprising a
gas of hadrons,  but is not inconsistent with such a step in evolution 
of the matter. The final particle abundances, measured in an experiment, 
are obtained after all unstable hadronic resonances `$j$'
are allowed to disintegrate, contributing to the yields of
stable hadrons.
The unnormalized particle multiplicities are obtained 
differentiating \req{4abis} with respect to particle 
fugacity. 

Are we sure that all  particle yields produced predominantly 
in the statistical hadronization process?
Already in 1982, we were pointing out that
rarely produced particles such as $\Omega, \overline\Omega$ 
would be the first to
show additional enhancement originating in either new
mechanisms, or simply in the fact that strangeness correlation
would be favoring strange clusters which
alter statistical hadronization yields of 
$\Omega,\overline\Omega$ \bcite{Raf82a}. 

It is important that the reader understands the method how one
finds such exceptions from the statistical hadronization pattern. 
Given that we have many available particle yields, we can look
at the data fits omitting one result at a time, in order to see that
the fit is stable,  and that it does not depend on one single
experimental result.
In order to report statistical significance,  this procedure 
must be carried out,  it is not implemented in fit programs.

The omission of the  $\Omega$ yields (and to a lesser extend $\overline\Omega$)
in the study of SPS results at $\sqrt{s_{\rm NN}}=17.2$ GeV, 
had the effect of allowing for a much better convergence 
of the statistical hadronization fit. This means that
within the reported  $\Omega$ (and $\overline\Omega$)
particle yields there are additional contributions 
outside of the statistical hadronization picture. Subsequent 
study of the $m_\bot$-spectra of $\Omega+\overline\Omega$
have in fact confirmed that there is an additional soft
momentum component \bcite{Tor01}. There are several 
explanations today of this behavior. Whatever is the true 
mix of production mechanisms,
we simply do not include the $\Omega,\overline\Omega$
yields in the statistical hadronization analysis, keeping in mind that
their large yield is a challenge for cascade model simulations.

\section{Hadron data fits}      
We turn rightaway to discuss  the  RHIC data
obtained in $\sqrt{s_{\rm NN}}=130$ GeV run. 
The experimental results we consider are the 
(combined where possible) data of STAR, PHENIX, BRAHMS, PHOBOS, for more
discussion of the data origin, see \bcite{Bra01}. For comparison purposes,
we adopted the results nearly as presented in this reference.
However,  we did not fit natural results
such as $\pi^+/\pi^-=1$. We also do not use the results for
$\rm K^*, \bar K^*$ since these yields depend on the degree of rescattering of
resonance decay products. 
We assume, in our fits in \rt{RHIChad}, that the multistrange
weak interaction cascading $\Xi\to \Lambda$, in the STAR
result we consider \bcite{Cas2002},  is  cut by  vertex discrimination
and we use these yields without a further correction.

The data analysed are at the central rapidity region where, due to approximate  
longitudinal scaling, the effects of flow cancel and 
we can evaluate the full phase space yields
in order to obtain particle ratios.
In the last column in  \rt{RHIChad},
the chemical equilibrium fit, the large  $\chi^2$
originates in the inability to account for multistrange 
$\overline\Xi,\,\Xi$. Similar results as here shown \cite{Raf01},  are 
presented in Ref. \bcite{Bra01}, 
which work was done before the multistrange hadron
yields were reported.

\begin{table}[t]
\captt{RHIChad}{RHIC 130 analysis}{ 
\small
Fits of central rapidity hadron ratios for RHIC  $\sqrt{s_{\rm NN}}=130$ GeV run.
Top section: experimental results, followed by chemical parameters, 
physical property of the phase space (energy $E$ and entropy $S$ per baryon number b), 
and the fit error. Columns: data, full non-equilibrium
fit, nonequilibrium fit constrained by strangeness conservation and supersaturation 
of pion phase space,  and in the last column, equilibrium fit constrained by 
strangeness conservation, upper index $^*$ indicates quantities fixed by these 
considerations.
}
\vspace*{0.cm}
\begin{center}
\begin{tabular}{lcccc}
\hline\hline
                                        & Data  & Fit& Fit         & Fit$^{\rm eq}$   \\
                                        &       &    & $\rm s-\bar s=0$&  $\rm s-\bar s=0$\\
\hline
$ \rm{\bar p}/p$                            &0.64\ $\pm$0.07\ & 0.637 & 0.640 &  0.587 \\
$\rm{\bar p}/h^-$                          &                & 0.068 & 0.068 &  0.075 \\
${\overline\Lambda}/{\Lambda}$          &0.77\ $\pm$0.07\ & 0.719 & 0.718 &  0.679  \\
$\rm{\Lambda}/{h^-}$                       & 0.059$\pm$0.001 & 0.059 & 0.059 &  0.059  \\
$\rm{\overline\Lambda}/{h^-}$              & 0.042$\pm$0.001 & 0.042 & 0.042 &  0.040  \\
${\overline{\Xi}}/{\Xi}$                &0.83\ $\pm$0.08\ & 0.817 & 0.813 &  0.790  \\
${\Xi^-}/{\Lambda}$                     & 0.195$\pm$0.015 & 0.176 & 0.176 &  0.130  \\
${\overline{\Xi^-}}/{\overline\Lambda}$ & 0.210$\pm$0.015 & 0.200 & 0.200 &  0.152  \\
$\rm{K^-}/{K^+}$                           &0.88\ $\pm$0.05\ & 0.896 & 0.900 &  0.891  \\
$\rm{K^-}/{\pi^-}$                         & 0.149$\pm$0.020 & 0.152 & 0.152 &  0.145  \\
$\rm{K_S}/{h^-}$                           & 0.130$\pm$0.001 & 0.130 & 0.130 &  0.124  \\
${\Omega}/{\Xi^-}$                      &                & 0.222 & 0.223 &  0.208 \\
${\overline{\Omega}}/{\overline{\Xi^-}}$&                & 0.257 & 0.256 &  0.247 \\
${\overline{\Omega}}/{\Omega}$          &                & 0.943 & 0.934 &  0.935    \\
\hline
$T$                                       &           &158$\pm$ 1   & 158$\pm$ 1     &  183$\pm$ 1      \\
$\gamma_{\rm q}$                            &      &1.55$\pm$0.01 & 1.58$\pm$0.08  &  1$^*$    \\
$\lambda_{\rm q}$                           &      &1.082$\pm$0.010& 1.081$\pm$0.006& 1.097$\pm$0.006      \\
$\gamma_{\rm s}$                            &       & 2.09$\pm$0.03 &  2.1$\pm$0.1   &  1$^*$    \\
$\lambda_{\rm s}$                           &     &$\!$1.0097$\pm$0.015$\!$&  1.0114$^*$     & 1.011$^*$ \\
\hline
$E/b$[{\small GeV}]\ \                        &   &  24.6& 24.7  &  21    \\
$s/b$                                         &   &  6.1 &  6.2  &  4.2  \\
$S/b$                                         &   & 151  &  152  &  131  \\
$E/S$[{\small MeV}]\ \                        &   & 163  &  163  &  159  \\
\hline
$\chi^2/$dof                                 &   & 2.95/($10\!-\!5$)  &2.96$\!$/$\!$($10\!-\!4$)  & 73/($10\!-\!2$)  \\
\hline\hline
\vspace*{-0.9cm}
\end{tabular}
\end{center}
\end{table}

The chemical equilibrium fit yields $E/S=159\,\mbox{MeV}<T=183$ MeV,
and thus in order to satisfy the constraints arising from 
the Gibbs-Duham relation:
\beql{ESrelPAL}
{E\over S}+{PV\over S}=T+\delta T\,,\quad 
\delta T=\mu_{\rm b}\frac{{\rm b}}{S}\,,\quad  
\mu_{\rm b}=3T\ln \lambda_{\rm q}\,,
\eeq
the emission volume must be unrealistically large, since the pressure
is much smaller than the energy density $E/V$. This inconsistency is
to best of my knowledge 
well known to the promoters of the equilibrium hadronization  model,
yet they do nothing to resolve it. 

On the other hand, in the chemical nonequilibrium fits 
with $\gamma_{\rm s},\gamma_{\rm q}>1$
there is good agreement with  experimental data.
Results are as would be expected for a rapidly 
expanding fireball,  they  converge to the maximum
allowed value:
\beql{gamq}
\gamma_{\rm q}=\gamma_{\rm q}^{\rm c}=e^{m_\pi/2T_{\rm f}}\,,
\eeq
which corresponds to the maximum entropy in the pion gas.
The large values of $\gamma_{\rm q}>1$ 
is consistent with the need to hadronize the excess entropy of the 
QGP.  The volume problem we addressed above is absent since
$E/S=163>T=158$ MeV.
The value of the hadronization temperature $T=158$ MeV is 
 below the  central expected equilibrium phase transition
temperature, and this  hadronization  temperatures at RHIC is  
consistent with sudden breakup of a supercooled QGP fireball.
We conclude that as has been predicted 15 years ago the inclusion of the  
yields of multistrange antibaryons in the RHIC data analysis, 
along with allowance for chemical non-equilibrium ($\gamma\ne 1$), 
has the sensitivity to discriminate the different 
reaction scenarios. 

The  value of the thermal energy content $E/b=25$ GeV, 
seen in \rt{RHIChad}, is in very good 
agreement with expectations once we allow for the kinetic energy content
associated with longitudinal and transverse motion. The energy of each particle 
is `boosted' with the factor $\gamma_\bot^v\cosh y_\parallel$. 
For $v_\bot=c/\sqrt{3}$, we have $\gamma_\bot^v=1.22$.
The longitudinal flow range is about $\pm 2.3$ rapidity units, 
according to PHOBOS results.
To obtain the energy increase due to longitudinal flow, we have to multiply 
by the average,  
$\int dy_\parallel \cosh y_\parallel/y_\parallel\to\sinh(2.3)/2.3=2.15$, 
for a total average increase in energy by 
factor 2.62, which takes the full energy content to 
$E^v/b\simeq 65$ GeV as expected. This consistency reassures that
we have a physically relevant fit of the data, which respects 
overall energy conservation.

The strangeness content, $s/b=6$,  is slightly below 
the expected equilibrated QGP phase space, which would have yielded 
 8.6 strange quark pairs per baryon
at $\lambda_{\rm q}=1.08$. Thus
$\gamma_{\rm s}^{\rm QGP}=6/8.6\simeq 0.7$\,, which is greater than the value 
0.5 a similar discussion yields for  the SPS energy range. Using the
fitted value  $\gamma_{\rm s}^{\rm HG}=2.1$, we find at RHIC like at SPS, 
$\gamma_{\rm s}^{\rm HG}/\gamma_{\rm s}^{\rm QGP}\simeq 3$, which is 
the correct phase space size ratio for HG and QGP at these conditions.

\section{Final remarks}      
There is ample evidence, looking at particle spectra at RHIC,
that the evolution of the system is very rapid. We have 
pioneered this reaction model in the analysis of the 
SPS data. We have in particular pointed  out that
the explosive disintegration of the QGP
accompanied by supercooling below equilibrium 
phase transition temperature is a natural 
phenomenon \bcite{Raf00}. Thus, in the hadronization model, 
we study this is the natural background against which
we make decisions how to proceed and how to interpret the 
results obtained in analysis of particle spectra.

In sudden hadronization,  
$V^{\mathrm{HG}} \simeq V^{\mathrm{QGP}} $, 
the growth of volume is negligible,  
$T^{\mathrm{QGP}}\simeq T^{\mathrm{HG}}$, the temperature 
is maintained across the hadronization front, and the chemical occupancy 
factors in both states of matter accommodate the different magnitude of the particle
phase space. In this case, the QGP  strangeness when `squeezed' into the
smaller HG phase space results in 
${\gamma_{\rm s}^{\mathrm{HG}}/\gamma_{\rm s}^{\mathrm{QGP}}}\simeq 3$\,,
which is of the same magnitude as the unfrozen color degeneracy. 
This  theoretical expectation is observed both at the top SPS and at the 
RHIC-130 run. 

For the top SPS energy range, this interesting result could be ignored, since
accidentally  $\gamma_{\rm s}^{\mathrm{HG}} \simeq  \gamma_{\rm q}^{\mathrm{HG}}$.
For this reason  hadronization at SPS energy can be also modeled in terms of an
equilibrium hadronization model: The pion enhancement associated with 
the high entropy phase can be accommodated by use of two temperatures, one
for the determination of absolute particle yields, and another for 
determination of the spectral shape. Such an approach has
 similar number of parameters, as the method presented here, the only
inconsistency (with HBT) is the large volume required.

However,   already at RHIC-130 the
hadron phase space occupancy for strangeness is significantly larger than
for light quarks, see  \rt{RHIChad}.  It is the inclusion of the  
yields of multistrange antibaryons in the RHIC data analysis, which
leads to clear resolution of the hadronization mechanism.
The specific per baryon strangeness yield at RHIC is an order of
magnitude greater than at SPS.

We see at RHIC
considerable convergence of the hadron production 
around properties of suddenly hadronizing entropy and
strangeness rich  QGP. There is overwhelming
theoretical and experimental evidence that the chemical
non-equilibrium  yields of hadrons are required to understand
with the necessary precision the nature of the quark--gluon fireball
formed in these interactions.

\subsection*{Acknowledgments}
I would like to thank Horst St\"ocker for the organization of
this stimulating discussion session at 
SQM2001 in Frankfurt, and the invitation to present these remarks. 
Work supported in part by a grant from the U.S. Department of
Energy,  DE-FG03-95ER40937. Laboratoire de Physique Th\'eorique 
et Hautes Energies, University Paris 6 and 7, is supported 
by CNRS as Unit\'e Mixte de Recherche, UMR7589.


\end{document}